\pdfoutput=1
\documentclass[aps,prl,reprint,preprintnumbers,superscriptaddress,nofootinbib]{revtex4-1}
\usepackage{amsmath,amsfonts,amssymb,hyperref,color,graphicx,xspace, enumitem,slashed}
\usepackage[T1]{fontenc}

\usepackage{amsmath,amsthm,amssymb,multirow,psfrag}
\usepackage{epsfig}
\usepackage{color}

\definecolor{nicered}{rgb}{.7,.1,.1}
\definecolor{nicegreen}{rgb}{.1,.5,.1}
\definecolor{darkblue}{rgb}{0,0,.5}
\hypersetup{colorlinks, citecolor=nicegreen, linkcolor=nicered, urlcolor=darkblue}

\def\lsim{\mathrel{\rlap{\lower4pt\hbox{\hskip1pt$\sim$}}
  \raise1pt\hbox{$<$}}}
\def\gsim{\mathrel{\rlap{\lower4pt\hbox{\hskip1pt$\sim$}}
  \raise1pt\hbox{$>$}}}
\newcommand{\vev}[1]{ \left\langle {#1} \right\rangle }
\newcommand{\bra}[1]{ \langle {#1} | }
\newcommand{\ket}[1]{ | {#1} \rangle }

\newcommand{\beq}{\begin{equation}}
\newcommand{\eeq}{\end{equation}}
\newcommand{\bea}{\begin{eqnarray}}
\newcommand{\eea}{\end{eqnarray}}
\newcommand{\nn}{\nonumber\\}

\newcommand{\bwt}{\begin{widetext}}
\newcommand{\ewt}{\end{widetext}}

\newcommand{\mbphi}{\overline{m}_\phi}
\newcommand{\mbpsi}{\overline{m}_\psi}

\newcommand{\fref}[1]{Fig.\,\ref{fig:#1}} 
\newcommand{\eref}[1]{Eq.\,\eqref{Eq:#1}}

\begin{document}

\title{Schwinger Current in de Sitter Space}

\author{Mar Bastero-Gil}
\email{mbg@ugr.es}
\affiliation{Departamento de F\'{i}sica Te\'{o}rica y del Cosmos and CAFPE,\\
Universidad de Granada, Campus de Fuentenueva, E-18071 Granada, Spain}

\author{Paulo B. Ferraz}
\email{paulo.ferraz@student.uc.pt}
\affiliation{Departamento de F\'{i}sica Te\'{o}rica y del Cosmos and CAFPE,\\
Universidad de Granada, Campus de Fuentenueva, E-18071 Granada, Spain}
\affiliation{Universidad de Coimbra, Faculdade de Ciências e Tecnologia da Universidade\\
de Coimbra and CFisUC, Rua Larga, 3004-516 Coimbra, Portugal}

\author{António Torres Manso}
\email{atmanso@uc.pt}
\affiliation{Universidad de Coimbra, Faculdade de Ciências e Tecnologia da Universidade\\
de Coimbra and CFisUC, Rua Larga, 3004-516 Coimbra, Portugal}

\author{\\Lorenzo Ubaldi}
\email{lorenzo.ubaldi@ijs.si}
\affiliation{Jo\v{z}ef Stefan Institute, Jamova 39, 1000 Ljubljana, Slovenia}
\affiliation{Institute for Fundamental Physics of the Universe (IFPU), \\ Via Beirut 2, 34014 Trieste, Italy}

\author{Roberto Vega-Morales}
\email{rvegamorales@ugr.es}
\affiliation{Departamento de F\'{i}sica Te\'{o}rica y del Cosmos and CAFPE,\\
Universidad de Granada, Campus de Fuentenueva, E-18071 Granada, Spain}

\date{\today}

\begin{abstract}
We study classical background electric fields and the Schwinger effect in de Sitter space.~We show that having a constant electric field in de Sitter requires the photon to have a tachyonic mass proportional to the Hubble scale.~This has physical implications for the induced Schwinger current which affect its IR behaviour.~To study this we recompute the Schwinger current in de Sitter space for charged fermions and minimally coupled scalars imposing a physically consistent renormalization condition.~We find a \emph{finite and positive} Schwinger current even in the massless limit.~This is in contrast to previous calculations in the literature which found a negative IR divergence.~We also obtain the first result of the Schwinger current for a non-minimally coupled scalar, including for a conformally coupled scalar which we find has very similar behaviour to the fermion current.~Our results may have physical implications for both magnetogenesis and inflationary dark matter production.
\end{abstract}

\maketitle

\begin{center}
{\bf \textsc{Introduction}}
\end{center}

A strong electric field is expected to produce electron-positron pairs out of the quantum field theory (QFT) vacuum.~This phenomenon was first hinted at in 1936 by Euler and Heisenberg~\cite{Heisenberg:1936nmg}, remarkably long before our modern theory of Quantum Electrodynamics (QED), then firmly predicted in a seminal paper by Schwinger~\cite{Schwinger:1951nm}.~Today the effect goes under the name of Schwinger pair production, but it has yet to be observed in the laboratory\footnote{Some condensed matter experiments have reported observation of the Schwinger effect~\cite{Berdyugin:2021njg,Schmitt:2022pkd} utilizing graphene as an analog of the QFT vacuum~\cite{Allor:2007ei} where electron-hole pairs replace $e^+e^-$ pairs.}.~This puts us in the fascinating situation where one of the fundamental predictions of QED, possibly our most successful theory of nature, has yet to be tested experimentally.~The primary reason is that one needs an intense electric field, of the order of the electron mass squared.~This is still beyond current technology, but there is hope to observe this phenomenon in a laboratory in the not too distant future~\cite{Dunne:2008kc,Fedotov:2022ely}.

These technological challenges on Earth could potentially be overcome in the early Universe during inflation, where there is enormous amounts of  energy available to generate strong electric fields.~Furthermore, and as we examine in detail, in an expanding spacetime, the Schwinger effect can occur even with very weak electric fields.~Mechanisms for generating the necessary constant electric field during inflation can be easily constructed by coupling a $U(1)$ gauge field to the expanding spacetime~\cite{Anber:2006xt,Anber:2009ua,Graham:2015rva,Adshead:2016iae,Bastero-Gil:2018uel,Bastero-Gil:2021wsf}.~This raises the interesting possibility that the first experimental observation of the Schwinger effect could come from cosmological signals~\cite{Chua:2018dqh}.

The Schwinger current in de Sitter space was first computed in $1+1$ dimensions~\cite{Frob:2014zka} and then in $3+1$ for minimally coupled spin 0~\cite{Kobayashi:2014zza,Hayashinaka:2016dnt,Banyeres:2018aax} and spin 1/2~\cite{Hayashinaka:2016qqn} charged particles.~The current is formally divergent, but remarkably, can be obtained analytically (non-perturbatively) and made finite utilizing an appropriate renormalization procedure.~In $3+1$ dimensions both the scalar and fermion currents were found to have a peculiar negative IR divergence in the small electron\,\footnote{For the remainder of this paper we use the label ``electron" for both charged scalars and fermions interchangeably.} mass limit.~This implies not only a divergent current, but one which flows in the opposite direction as the electric field.~Understanding this seemingly unphysical behavior is not just an academic curiosity, it can have important implications in primordial magnetogenesis~\cite{Kobayashi:2014zza} and play a role in the production of dark matter~\cite{Arvanitaki:2021qlj,Bastero-Gil:2023htv,Bastero-Gil:2023mxm}.

In this work we argue that the appearance of these IR divergences is due to the renormalization procedures implicitly utilized in~\cite{Kobayashi:2014zza,Banyeres:2018aax,Hayashinaka:2016dnt}.~Imposing physical renormalization conditions which are consistent with a constant electric field in de Sitter space, we show that the renormalized current is {\em free of UV and IR divergences}, both for scalars and fermions.~In deciding how to fix the renormalization conditions, we first point out that in order to sustain a constant electric field in de Sitter space it is unavoidable to have a tachyonic instability.~Requiring the electric field to be constant and uniform both inside and outside the Hubble horizon, as was assumed in~\cite{Kobayashi:2014zza,Banyeres:2018aax,Hayashinaka:2016dnt}, implies that the photon, treated as a classical \emph{dynamical} field, must have a tachyonic mass $m_A^2 = -2H^2$, where $H$ is the Hubble constant.~As we will show, given the tachyonic mass condition, this then leads to a modified on-shell renormalization condition which gives an IR finite renormalized current.~More details of our analysis can be found in an accompanying paper~\cite{Bastero-Gil:2025nfi}.

%%%%%%%%%%%%%%
%%%%%%%%%%%%%%
%%%%%%%%%%%%%%
\begin{center}
{\bf \textsc{Classical constant electric field\\ in de Sitter}}
\end{center}

Here, as in previous studies of the Schwinger current~\cite{Kobayashi:2014zza, Hayashinaka:2016qqn, Banyeres:2018aax}, we consider a classical constant electric field in de Sitter space which is already present during inflation and treat the geometry of spacetime as a background.~This means we do not deal with the dynamics of the energy-momentum tensor and do not worry about renormalizing vacuum energy.~Crucially, and in contrast to previous works, we treat the classical gauge field $A_\mu$, which is responsible for setting the constant electric field, as dynamical.~This is because the (only) divergence in the theory, associated with the photon field normalization, is encoded in the vacuum polarization diagram~\cite{Bastero-Gil:2025nfi}, consisting of an electron loop with two external photon legs.~To remove the divergence we need the counterterm associated with such a diagram.~This requires the two-point function of the photon field which is obtained by treating the field dynamically.

We begin the action for a free abelian gauge theory in a de Sitter spacetime background,
\bea
	S=-\int d^4 x \sqrt{-g}\,\frac{1}{4}F^{\mu\nu}F_{\mu\nu}\,.
\eea
Here the metric is $g_{\mu\nu} = a^2(\tau)(-{\rm d}\tau^2 + {\rm d}\vec x^2)$, with the conformal time given by $\tau$,
\begin{equation}
\tau = - \frac{1}{aH} < 0 \, , \qquad H = \frac{{\rm d}a}{a^2 {\rm d}\tau} = {\rm const.} \, ,
\end{equation}
and $F_{\mu\nu} = \partial_\mu A_\nu - \partial_\nu A_\mu$.~We are interested in the configuration which realizes a constant and uniform electric field.~This is the one measured by a comoving observer with 4-velocity $u^\mu$ ($u^i = 0$, $u_\mu u^\mu = -1$), given by $E_\mu = u^\nu F_{\mu\nu}$ and such that the field strength $E_\mu E^\mu = E^2 = {\rm const}$.~The gauge field configuration that produces such a constant electric field in de Sitter space in the $z$ direction is given by~\cite{Kobayashi:2014zza,Hayashinaka:2016qqn,Banyeres:2018aax},
 \bea\label{Eq:Edef}
 A_\mu=\frac{E}{H^2 \tau} \delta_\mu^z \, .
 \eea
Throughout this work, and as done in previous literature, we treat the gauge field as classical.~However, as discussed above and differently from previous literature, we treat it as dynamical, rather than as a background, requiring it to solve the equation of motion,
\bea \label{Eq:EOMA}
g^{\alpha \mu}\partial_\alpha F_{\mu\nu} = 0 \, .
\eea
It is easy to check that \eref{Edef} does not solve \eref{EOMA}:
\bea
g^{\alpha \mu}\partial_\alpha F_{\mu\nu} = - 2 H^2 \tau^2 \frac{E}{H^2 \tau^3} \delta_\nu^z = -2 H^2 A_\nu \neq 0 \, .
\eea
The easiest way to satisfy the equation of motion is to introduce a tachyonic mass,
\bea\label{Eq:Amass}
m_A^2 = -2H^2 \, ,
\eea
into the action of the gauge field,
\bea\label{Eq:LmA}
	S=-\int d^4 x \sqrt{-g}\,\left(\frac{1}{4}F^{\mu\nu}F_{\mu\nu}+\frac{1}{2} m_A^2 A_\mu A^\mu\right)\,.
\eea
The photon mass $m_A$ should be thought of as a Stueckelberg mass~\cite{Ruegg:2003ps}, obtained upon the replacement
$\frac{1}{2} m_A^2 A_\mu^2 \to \frac{1}{2} m_A^2 \left( A_\mu + \frac{1}{m_A} \partial_\mu \sigma \right)^2$, with $\sigma$ the Stueckelberg field, and the addition of the gauge fixing term ${\cal L}_{\rm gf} = - \sqrt{-g} \frac{1}{2\xi_A} \left( \partial_\mu A^\mu - \xi_A m_A \sigma \right)^2$.~The theory is then gauge invariant.~We see that in order to have a constant electric field in de Sitter space, 
we need a source term which breaks conformal invariance and provides a tachyonic condition, necessary to compensate for the exponential expansion of space.

A second possible solution is to instead introduce a modified kinetic term of the form $(\tau_e/\tau) F_{\mu\nu} F^{\mu\nu}$ where $\tau_e$ is a constant which could be fixed as the time at the end of inflation for example.~As we show in Appendix A of~\cite{Bastero-Gil:2025nfi}, this leads to the same equations of motion as introducing a tachyonic mass with a canonically normalized kinetic term.~In both scenarios the electric field originates from a transverse component of the gauge field, which we treat as purely classical, with no contribution from the longitudinal mode.~It should not be surprising that in de Sitter space the photon must have a tachyonic mass since, as space expands exponentially, the electric field must be continuously fed exponentially meaning that a tachyonic instability is required to maintain a constant electric field.~We will see below the crucial role the tachyonic condition $m_A^2 = -2H^2$ plays in renormalization and obtaining a physically consistent Schwinger current.

%%%%%%%%%%%%%%
%%%%%%%%%%%%%%
%%%%%%%%%%%%%%
\begin{center}
\noindent{\bf \textsc{Renormalized Lagrangian with constant electric field in de Sitter.}}
\end{center}

Since we are interested in computing the current of charged particles generated by the constant electric field out of the vacuum, one might worry about radiative effects which would require quantizing the photon field.~However, shortly after they are produced the charged particles reach terminal velocity~\cite{Bastero-Gil:2023mxm} implying that very quickly the current reaches a constant value.~We can therefore neglect radiative effects and solve the theory exactly quantizing only the electron field, further justifying our classical treatment of the electric field. 

We can write the renormalized Lagrangian in de Sitter for a classical massive photon coupled to a current,
\bea\label{Eq:renormlag}
\mathcal{L} &=& 
 - \frac{1}{4} (1 + \delta_3) (F_{\mu \nu})^2 - \frac{1}{2} m_A^2 A_\mu A^\mu 
 - A_\mu J^\mu + ... ~~.
\eea
All fields and parameters are understood to be renormalized, and the conserved current $J^\mu$, which depends on the gauge coupling $e$, can be for fermions or scalars.~One can show~\cite{McKeon:2006ym} that the renormalized photon mass $m_A$ is related to the bare one as $m_A^2 = (1- \delta_3)m_{A_0}^2$ in the same way the renormalized charge $e$ is related to the bare charge $e^2 = (1-\delta_3) e^2_0$, so there is no additional divergence.~The counterterm $\delta_3$ is used to cancel the log divergence in the photon field normalization.

\begin{figure}
\vspace{.4cm}
\centering 
\includegraphics[width=.8\columnwidth]{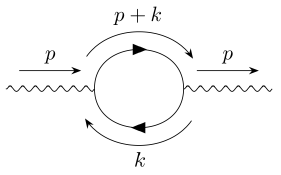}
\caption{Vacuum polarization diagram.}
\label{fig:selfenergy}
\end{figure}

In order to define our theory we must first fix the counterterm with an appropriate \emph{physical} renormalization condition.~Since the field $A_\mu$ is classical, the obvious choice is to pick on-shell renormalization conditions.~This requires the renormalized
two point function to have a pole at $p^2 = -m_A^2$ with residue $-i$ which fixes,
\bea
\Pi(p^2 = -m_A^2) = 0, 
\eea
where $e^2\Pi(p^2)$ is defined as the coefficient of $-i\left(p^2 g^{\mu \nu} - p^\mu p^\nu\right)$ in the sum of all 1PI contributions to the photon 2-point correlation function~\cite{Schwartz:2014sze}.~Since we have a classical electric field and $A_\mu$ is not quantized, $\Pi(p^2)$ is determined exactly at one loop, with the corresponding diagram shown in~\fref{selfenergy} (with a second diagram for scalars).~This renormalization condition leads to the condition on the counterterm,
\bea\label{Eq:del3fix}
\delta_3 = - e^2\Pi(-m_A^2) .
\eea
Obviously, for a massless photon one takes $m_A^2 = 0$ which holds in flat space.~As we discuss more below and show in detail in~\cite{Bastero-Gil:2025nfi}, this is equivalent to the renormalization procedures used in~\cite{Kobayashi:2014zza,Hayashinaka:2016qqn,Banyeres:2018aax}.~However, as we have discussed above, a massless photon is inconsistent with a constant classical electric field in de Sitter space which requires a tachyonic mass $m_A^2 = -2H^2$.~Thus we take the counterterm to be fixed by the condition,
\bea\label{Eq:del3fix2}
\delta_3 = - e^2\Pi(2H^2) .
\eea
Note that in contrast to renormalization conditions defined (implicitly) in previous calculations in the literature,\,\eref{del3fix2} is independent of considerations of the Schwinger current or renormalization.~As we'll see, with this renormalization condition, the renormalized Schwinger current in de Sitter space is not only UV finite, but crucially, it is also free of the negative IR divergences which were found in previous calculations for both scalars~\cite{Kobayashi:2014zza} and fermions~\cite{Hayashinaka:2016qqn}.~In~\cite{Banyeres:2018aax} the authors argued that the negative IR divergent current is not physical and correctly connected the origin of the issue with the renormalization condition, but did not connect such a condition to the tachyonic mass of the vector in de Sitter space, which is a crucial point of our work.

%%%%%%%%%%%%%%
%%%%%%%%%%%%%%
%%%%%%%%%%%%%%
\begin{center}
{\bf \textsc{Counterterm with a constant\\ electric field in de Sitter}}
\end{center}

When computing the vacuum polarization diagram we cannot calculate analytically the loop diagram in de Sitter space.~To obtain an analytic result, one must work in the Minkowski limit in both the spacetime integral and the electron propagator.~The vacuum polarization diagram is then given by the well known result from QED which depends on the invariant mass squared $(p^2)$ of the photon.~This properly takes care of the UV modes in the counterterm thus allowing divergences in the regularized current to be canceled ensuring a finite renormalized current.~However, if we follow the renormalization procedure used (implicitly) in previous calculations~\cite{Banyeres:2018aax} in the literature, assuming a massless photon and setting $p^2=0$, a negative IR divergence is introduced into the counterterm when the electron mass is taken to zero.~As we show in detail in~\cite{Bastero-Gil:2025nfi}, this negative IR divergence is then introduced into the physical \emph{renormalized} current as there is no corresponding IR divergence in the \emph{regularized} current, which is computed exactly, that could cancel it.

To see this explicitly we take the standard flat space result from scalar QED for the vacuum polarization~\cite{Schwartz:2014sze}.~Setting $p^2 = 0$, we have for the counterterm, 
\bea\label{Eq:d3phi1}
\delta_3 &=& -e^2\Pi(0) 
=
\left(\frac{e}{12\pi}\right)^2
\Big[ -3 \ln \Big(\frac{\Lambda ^2}{m^2} \Big) \Big]\, ,
\eea
with a factor 4 included in the case of a charged fermion.~Here $m$ is the mass of the electron which enters through the propagator (in the Minkowski limit) and $\Lambda$ is the mass of the Pauli-Villars fields used to regulate the UV divergence (see Appendix A in~\cite{Bastero-Gil:2025nfi} for details).~With this counterterm we can reproduce previous results in the literature for the renormalized current of a minimally coupled massive charged scalar~\cite{Kobayashi:2014zza,Banyeres:2018aax} and a massive charged fermion~\cite{Hayashinaka:2016qqn} which were obtained using different renormalization procedures.~While the UV log divergence in~\eref{d3phi1} cancels the one in the regularized current, we see explicitly the negative IR divergence in the $m \to 0$ limit.~However, as we have discussed above, the presence of a constant (dynamical) electric field in de Sitter space requires $-p^2 = m_A^2 = -2H ^2$, independently of renormalization or the Schwinger current.

For a charged scalar $\phi$, if we evaluate the vacuum polarization diagram with $p^2 = 2H^2$ instead we obtain~\cite{Bastero-Gil:2025nfi} in the limit $m_\phi / H \to 0$,
\bea\label{Eq:del3phiapplim}
\delta_3^{\phi}
&\approx&
\frac{e^2}{144\pi^2}
\Big[ -3 \ln \Big(\frac{\Lambda ^2}{H^2}\Big) - 5.9 \Big] ,
\eea
which we see is IR finite.~For a charged fermion loop in the $m_\psi/H \to 0$ limit we find~\cite{Bastero-Gil:2025nfi},
\bea\label{Eq:del3psiapplim}
\delta_3^{\psi}
&\approx&
\frac{e^2}{36\pi^2 }
\Big[
-3 \ln \Big(\frac{\Lambda ^2}{H^2}\Big)
- 2.9\Big]  ,
\eea
which again we see is IR finite.~While~\eref{del3phiapplim} and~\eref{del3psiapplim} are obtained in the Minkowski and large loop momentum limit, they are perfectly self consistent and free of IR divergences which have been cured, not by hand, but by imposing the tachyonic photon mass condition needed for sustaining a constant electric field in de Sitter space which is independent of considerations of the Schwinger current or renormalization.~As we discuss more below and show in detail in~\cite{Bastero-Gil:2025nfi}, with these counterterms one obtains a renormalized Schwinger current which is \emph{both UV and IR finite}.

%%%%%%%%%%%%%%
%%%%%%%%%%%%%%
%%%%%%%%%%%%%%
\begin{center}
{\bf \textsc{Renormalized Schwinger current\\ and modified counterterms}}
\end{center}

Until now, our previous discussion has been completely independent of considerations of the Schwinger current.~With the counterterm defined in~\eref{del3fix2} we can define the physical renormalized current.~Varying the action of \eref{renormlag} with respect to the gauge field,
\bea \label{Eq:currentdef} 
\left(1 + \delta _3 \right)\partial^{\mu }F_{\mu \nu } -  m_A^2 A_\nu = J_{\nu }  \, .
\eea 
Using the field configuration given in \eref{Edef}, we have $\partial^\mu F_{\mu\nu} = -2 a H E \delta_\nu^z$.~While we treat $A_\mu$ as classical, we are going to compute a current arising from the quantum fluctuations of charged scalar and fermion fields.~To compare such a current to the other classical terms in~\eref{currentdef} we should take its vacuum expectation value 
$\bra{0} J_\mu \ket{0} \equiv \vev{J_\mu}$ where $\ket{0}$ is the Bunch-Davies vacuum of the charged field giving for~\eref{currentdef},
\bea\label{Eq:rencurrent2}
\partial^{\mu }F_{\mu \nu } - m_A^2 A_\nu = \vev{J_\nu}_{\rm reg} + 2 aH E \delta_\nu^z \delta_3  \, .
\eea 
We have added the subscript `reg' to the expectation value of the current which is divergent and must therefore be regularized.~We emphasize that the same scheme must be used to regularize both $\vev{J_\mu}$ and $\delta_3$.~With an appropriate renormalization condition to fix the finite part of $\delta_3$ we can obtain an unambiguous renormalized current,
\bea \label{Eq:Jrendef}
 \langle J_z \rangle_{\rm ren} =  \langle J_z \rangle_{\rm reg} + 2 aHE \,\delta_3 \, .
\eea 
%\
Calculating $\langle J_z \rangle_{\rm reg}$ is involved and we refer the reader to~\cite{Kobayashi:2014zza,Hayashinaka:2016qqn,Banyeres:2018aax} for details.~We have repeated the calculation~\cite{Bastero-Gil:2025nfi} and confirm previous results for both scalars and fermions.~The result is non-perturbative containing all effects from de Sitter space and, crucially, there is no IR divergence in the small electron mass limit.

In contrast, the counterterms in~\eref{del3phiapplim} and~\eref{del3psiapplim} have been obtained taking the Minkowski and large loop momentum limits in the vacuum polarization diagram.~Thus, the only information they have about de Sitter space is through the tachyonic condition $p^2 = 2H^2$.~So although there is no IR divergence, we should not expect to properly capture all IR behavior.~We can see this in~\eref{del3phiapplim} and~\eref{del3psiapplim} which contain small (constant) negative finite parts that, in the small electron mass limit, lead to a constant, but slightly negative renormalized current~\cite{Bastero-Gil:2025nfi}.~This implies the unphysical behavior of a current flowing opposite to the electric field.

We expect that including corrections to the vacuum polarization from de Sitter space ensures a positive current, but in general, they cannot be included analytically.~One that can be included analytically is the curvature correction to the electron mass coming from a non-minimal coupling to gravity in the case of a scalar or the spin connection in the case of the charged fermion.~These corrections are automatically included in the \emph{regularized} current which is computed non-perturbatively in de Sitter space directly from the Lagrangian.

In the calculation of the counterterm the curvature corrections enter through the mass in the electron propagator.~To see this consider the equations of motion in a curved background for a free scalar $\phi$ with non-minimal coupling to gravity $\xi$ and a free fermion $\psi$,
\bea\label{Eq:phiEOM}
(\Box + m_\phi^2 + \xi R)\phi&=&0 \, ,\\
(\Box + m_\psi^2 + \frac{1}{4} R ) \psi &=& 0 .
\eea
The latter is obtained~\cite{Parker:2009uva} by taking the square of the Dirac operator, $\slashed D \slashed D \psi =0$ with $\slashed D = \gamma^\mu D_\mu$ and $D_\mu = \partial_\mu + B_\mu$ where $B_\mu$ is the spin connection.~One consequence of the curved background is the addition of a term proportional to $R$ in the Klein-Gordon equation.~In particular, in de Sitter space, $R=12H^2$ is a constant and such a term can be incorporated into the squared mass term.~Thus we can take the free scalar propagator proportional to $(k^2 + m_\phi^2 + \xi R)^{-1}$ and the free fermion propagator proportional to $(k^2 + m_\psi^2 + R/4)^{-1}$.~There are additional (time dependent) corrections from de Sitter space inside the $\Box$ operator which our calculation does not account for, but to our current knowledge they cannot be computed analytically.~Nevertheless, we expect that if one were eventually to perform the full calculation of the counterterm in de Sitter space, the resulting renormalized current would be positive in all regimes. 

With these modified propagators we repeat the calculation of $\delta_3$ in Minkowski space.~Imposing the renormalization condition in~\eref{del3fix2}, we obtain for the finite parts of the counterterms (see Appendix A in~\cite{Bastero-Gil:2025nfi} for details),
\begin{widetext}
\bea\label{Eq:delta3PV}
\delta_3^{\phi} = \left(\frac{e}{12\pi}\right)^2
\left[  3 \ln \left( \mbphi^2 + 12\xi \right)
-12 \left( \mbphi^2 + 12\xi \right) 
 + 6\left(2 \left( \mbphi^2 + 12\xi \right) 
+ 1\right)^{3/2}\coth ^{-1}
\left(\sqrt{2\left( \mbphi^2 + 12\xi \right)+1}\right)  -8 \right],~~~
\eea
\end{widetext}
for the scalar case, while for the fermion case we obtain,
\begin{widetext}
\bea\label{Eq:delta3PVf}
\delta_3^{\psi} = 
\frac{e^2}{36\pi^2 } &
\Big[
 3 \ln \left(\mbpsi^2+3\right) + 6 (\mbpsi^2 + 3)  
-6 \left((\mbpsi^2 + 3) - 1\right) 
\sqrt{2(\mbpsi^2+3)+1}\, \coth ^{-1}
\left({\sqrt{2(\mbpsi^2+3) + 1}}\right) - 5\Big],
\eea
\end{widetext}
where we have defined $\mbphi \equiv m_\phi/H, \mbpsi \equiv m_\psi/H$ and taken $R = 12H^2$ for the curvature in de Sitter space.~We see explicitly that the curvature corrections to the electron mass give additional positive finite contributions in the counterterm.~In the massless electron limit this leads to a \emph{positive} finite part for $\delta_3$ in contrast to~\eref{del3phiapplim} and~\eref{del3psiapplim} where the curvature correction to the electron mass is neglected and the finite part is negative.

%%%%%%%%%%%%%%
%%%%%%%%%%%%%%
%%%%%%%%%%%%%%
\begin{center}
{\bf \textsc{Behavior of Renormalized Current}}
\end{center}

With these counterterms we use~\eref{Jrendef} to obtain renormalized Schwinger currents for charged scalars and fermions which are UV and IR finite as well as positive over all regions of parameter space.~Explicit analytic expressions for the renormalized charged scalar and fermion currents are given in~\eref{Jphiren} and~\eref{Jpsiren} of the Appendix with details of the calculations given in~\cite{Bastero-Gil:2025nfi}.~In~\fref{currents} we show our results (solid) for the dimensionless Schwinger current $\mathcal{J} \equiv  \langle J_z \rangle_{\rm ren}/a e H^3$ as a function of $\lambda \equiv eE/H^2$ for both scalars and fermions and a few choices of parameters as indicated in the figure.~We see the current is always positive for both fermions and scalars (which we have multiplied by 2).~For a minimally coupled scalar ($\xi = 0$) we see the IR hyperconductivity behavior found in~\cite{Kobayashi:2014zza,Banyeres:2018aax}.~We also see that the current for a conformally coupled scalar ($\xi = 1/6)$, which we have obtained here for the first time, has the same behavior as the fermion current.~A detailed analysis of the behavior of the current in different limits can be found in~\cite{Bastero-Gil:2025nfi}.
\begin{figure}
\vspace{.4cm}
\centering
\includegraphics[width=\columnwidth]{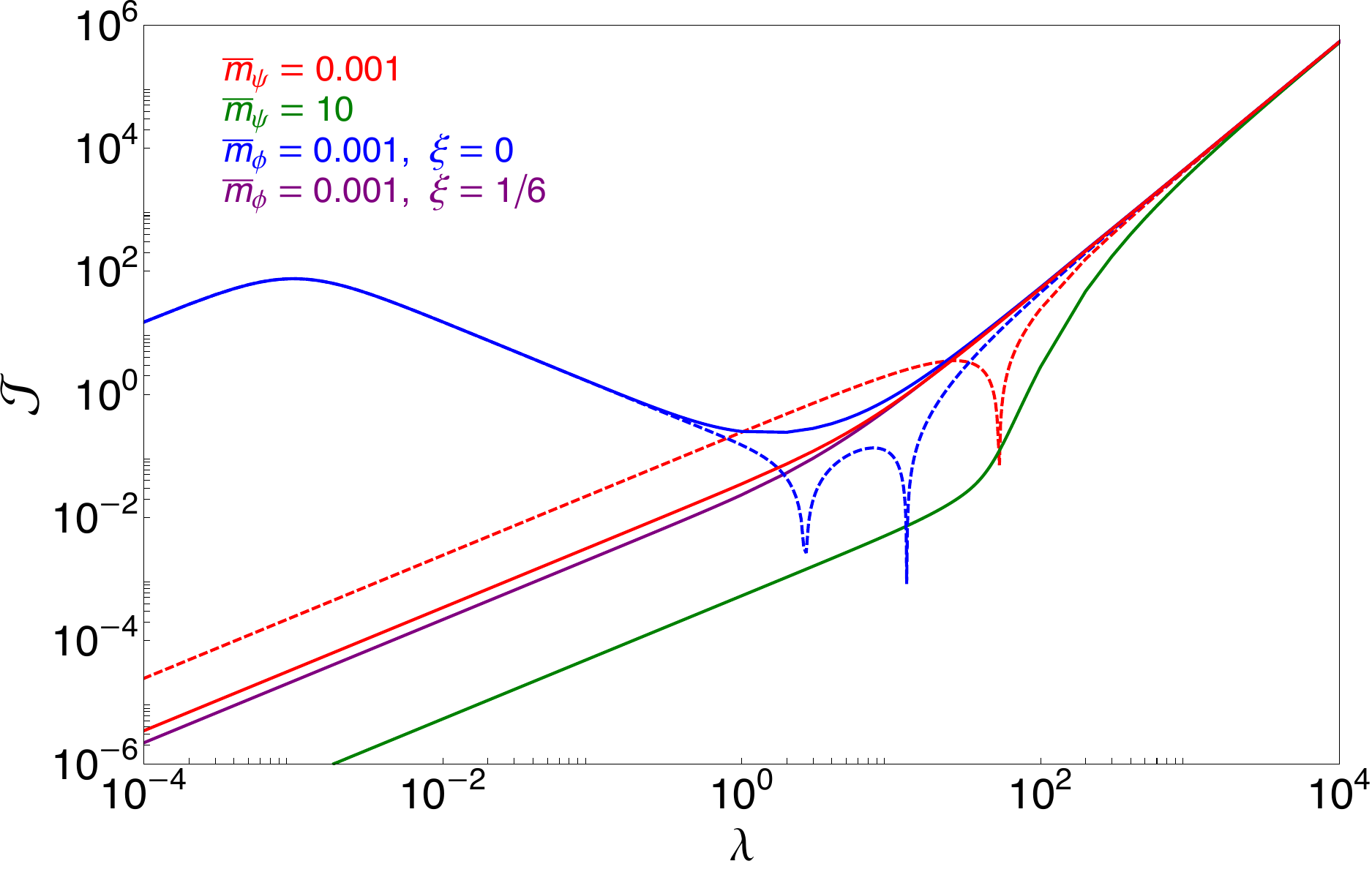}
\caption{The renormalized (dimensionless) Schwinger current $\mathcal{J} \equiv  \langle J_z \rangle_{\rm ren}/a e H^3$ for a charged fermion ($\psi$) and scalar ($\phi$) (multiplied by 2) as a function of $\lambda \equiv eE/H^2$ (solid).~Explicit analytic expressions are given in~\eref{Jphiren} and~\eref{Jpsiren} of the Appendix.~We also show for comparison, results for the Schwinger renormalized current obtained previously (dashed) in the literature~\cite{Kobayashi:2014zza,Hayashinaka:2016qqn}.~Between the cusps the scalar current is negative while the fermion current is negative to the left of the cusp so we have plotted the absolute value for each.
\vspace{-4ex}}
\label{fig:currents}
\end{figure}

We also show for comparison the currents obtained in previous results (dashed) in the literature~\cite{Kobayashi:2014zza,Hayashinaka:2016qqn} which contain an IR divergent $\ln \left(m^2/H^2 \right)$ term arising due to implicitly assuming a massless photon in de Sitter space, as discussed above.~Between the cusps the scalar current is negative while the fermion current is negative to the left of the cusp so we plot the absolute value.

\begin{center}
{\bf \textsc{Summary and Discussion}}
\end{center}

We have revisited classical background electric fields and the Schwinger effect in de Sitter space which can have important implications for the study of primordial magnetogenesis, inflationary dark matter production, and possibly other interesting cosmological mechanisms.~We first pointed out that a constant electric field in de Sitter space requires the photon to have a tachyonic mass (see~\eref{Amass}).~We then showed how this fixes the counterterm (see~\eref{del3fix2}) needed to absorb the single divergence present in the theory.~With this counterterm we then recomputed the renormalized Schwinger currents in de Sitter for both charged fermions and minimally coupled scalars finding them to be \emph{UV and IR finite} as well as \emph{positive} for all values of parameters.~In particular, our results are free of the peculiar negative IR divergence found in previous calculations in the literature~\cite{Kobayashi:2014zza,Hayashinaka:2016qqn}.~We have traced the origin of this IR divergence to renormalization conditions which implicitly assumed a massless photon instead of tachyonic.~We have also computed for the first time the Schwinger current in de Sitter space for a conformally coupled scalar also finding it to be positive and finite with very similar behavior to the charged fermion current.~Additional details and discussion of our analysis can be found in an accompanying paper~\cite{Bastero-Gil:2025nfi}.

We emphasize that since the gauge field is not quantized (only the charged fields), the one-loop calculation to fix the counterterm contains the full quantum information so our results are non-perturbative.~The calculation of the \emph{regularized} current is also non-perturbative and done in de Sitter space.~However, since the calculation of the counterterm is done in Minkowski spacetime, with the inclusion of curvature corrections
to the electron mass and imposing the tachyonic photon mass condition, our final result for the \emph{renormalized} Schwinger current is non-perturbative, but not exact.~It can in principle be improved with a calculation of the counterterm containing the full information of de Sitter space.~Nevertheless, our results resolve the puzzling IR behavior found in previous literature and are an important step in understanding the Schwinger current in de Sitter space.

\vspace{0.6cm}
%%%%%%%%%%%%%%
%%%%%%%%%%%%%%
%%%%%%%%%%%%%%
{\bf \textsc{Acknowledgements:}}\,The authors thank Pedro Garcia Osorio, Manel Masip, Jose Santiago and Takeshi Kobayashi for useful comments and discussions.\,This work has been partially supported by Junta de Andaluc\'ia Projects\, P21-00199, A-FQM-472-UGR20 (fondos FEDER) and by SRA (10.13039/501100011033) and ERDF under grant PID2022-139466NB-C21 (R.V.M.) as well as by PID2022-140831NB-I00 funded by MICIU/AEI/10.13039/501100011033 and FEDER,UE (M.B.G.,A.T.M.),\,FCT CERN grant 10.54499/2024.00252.CERN\,(M.B.G.,A.T.M.,P.F),\,and FCT Grant No.SFRH/BD/151475/2021and 10.54499/SFRH/BD/151475/2021\,(P.F.).\,LU is supported by the Slovenian Research Agency under the research core funding No.P1-0035, and by the research grants J1-60026 and J1-4389.\,This article is based upon work from COST Action COSMIC WISPers CA21106, supported by COST (European Cooperation in Science and Technology).\,RVM thanks the Mainz Institute for Theoretical Physics (MITP) of the Cluster of Excellence PRISMA$^+$ (Project ID 390831469), for its hospitality and partial support during the completion of this work. 

\vspace{1cm}
\begin{center}
{\bf \textsc{Appendix}}
\end{center}

Here we give explicit expressions for the physical renormalized Schwinger currents for both non-minimally coupled charged scalars ($\phi$) and fermions ($\psi$).~Details of the calculation and analysis can be found in~\cite{Bastero-Gil:2025nfi} to which we refer the reader.~Defining the dimensionless ratios $\lambda \equiv eE/H^2$ and $\mbphi \equiv m_\phi/H$ we find for the renormalized charged scalar Schwinger current,
\begin{widetext}
\bea\label{Eq:Jphiren}
	\langle J_z^\phi \rangle_{\rm ren}
	&=& \langle J^\phi_z \rangle_{\mathrm{reg}} 
+ (2aH E)\delta^\phi_3\nonumber \\
	&=& a e H^3 \frac{\lambda}{4 \pi^2}  
	\Big[
	- \frac{2 \lambda^2}{15}
	+ F_\phi(\lambda, \mu)
	+ \frac{1}{6}\ln{ ( \mbphi^2 + 12\xi ) }
	- \frac{2}{3}( \mbphi^2 + 12\xi ) \nonumber\\ 
	&+& \frac{1}{3}\left(2 ( \mbphi^2 + 12\xi ) + 1\right)^{3/2}
	\coth ^{-1}\left(\sqrt{2 ( \mbphi^2 + 12\xi )  +1}\right) - \frac{4}{9} \Big] ,\nonumber\\
	~\nonumber\\
F_\phi(\lambda, \mu) &\equiv&  
\frac{45+4 \pi^2\left(-2+3 \lambda^2+2 \mu^2\right)}{12 \pi^3} \frac{\mu \cosh (2 \pi \lambda)}{\lambda^2 \sin (2 \pi \mu)}
- \frac{45+8 \pi^2\left(-1+9 \lambda^2+\mu^2\right)}{24 \pi^4} \frac{\mu \sinh (2 \pi \lambda)}{\lambda^3 \sin (2 \pi \mu)} \\
&+& 
\operatorname{Re}\Big[\int_{-1}^1 d r \frac{i}{16 \sin (2 \pi \mu)}
\left(-1+4 \mu^2+\left(7+12 \lambda^2-12 \mu^2\right) r^2-20 \lambda^2 r^4\right) \nonumber\\
&\times&
\left(\left(\mathrm{e}^{-2 \pi r \lambda}+\mathrm{e}^{2 \pi i \mu}\right) \psi\left(\frac{1}{2}+\mu-i r \lambda\right)
- \left(\mathrm{e}^{-2 \pi r \lambda}+\mathrm{e}^{-2 \pi i \mu}\right) 
\psi\left(\frac{1}{2}-\mu-i r \lambda\right)\right)\Big] ,\nonumber
\eea
\end{widetext}
where we have defined $\mu^2=\frac{9}{4} - (\mbphi^2 + 12\xi) - \lambda^2$ and $\xi$ is the non-minimal coupling to gravity.~Defining the dimensionless ratio $\mbpsi \equiv m_\psi/H$, we find for the renormalized charged fermion Schwinger current,
\begin{widetext}
\bea
\label{Eq:Jpsiren}
\langle J_{z }^\psi\rangle_{\rm ren} &=& 
\langle J_{z }^\psi\rangle_{\rm reg} 
+ (2a H E) \delta^\psi_3 \nonumber \\
&=& eaH^3 \frac{\lambda}{2\pi^2}
\Big[\frac{1}{2} + \frac{2 \lambda ^2}{15}
+ \frac{3 \mbpsi^2}{2 \lambda ^2} 
\Big(1+\frac{x}{2 \lambda } 
\ln (\frac{x -\lambda }{\lambda + x})\Big) 
+ F_\psi(\lambda,\mbpsi) 
+  \frac{1}{3} \ln (\mbpsi^2 + 3)  \nn
&+& \frac{2}{3}(\mbpsi^2 + 3)
- \frac{2}{3} \left( (\mbpsi^2 + 3) - 1\right) \sqrt{2(\mbpsi^2 + 3) + 1} \coth ^{-1}
\Big(\sqrt{2 (\mbpsi^2+3)+1}\Big) -\frac{5}{9}  \Big] ,  \\
~\nonumber\\
F_\psi(\lambda,\mbpsi) &\equiv& 
\frac{x\text{csch}(2\pi x)}{12\pi^3\lambda^2}\Big\{(45-\pi^2(11-12\lambda^2+8x^2))\cosh(2\pi \lambda)
-(45-\pi^2(11-72\lambda^2+8x^2))\frac{\sinh(2\pi \lambda)}{2\pi \lambda}\Big\}\nonumber\\
&-&\frac{\text{csch}(2\pi x)}{4}\operatorname{Re}
\Big[\int_{-1}^1dy(1+x^2 - (1+3\lambda^2 + 3x^2)y^2+5\lambda^2y^4)
\times\sum_{s=\pm}s(e^{2\pi \lambda y} - e^{-2\pi sx})\psi(i(\lambda y + sx))\Big]\nonumber\\
&-&\frac{3xM^2\text{csch}(2\pi x)}{8\lambda^3}\sum_{s=\pm}se^{-2\pi xs}(\text{Ei}(2\pi s(x+\lambda))-\text{Ei}(2\pi s(x-\lambda))), \nonumber
\eea
\end{widetext}
where we have also defined $x \equiv \sqrt{\mbpsi^2+\lambda^2}$.
%%%%%%%%%%%%%%
%%%%%%%%%%%%%%
%%%%%%%%%%%%%%
\bibliographystyle{apsrev4-2}
\bibliography{SchwingerCurrent}

\end{document}